\documentclass{aa}  
\usepackage{graphicx}
\usepackage{txfonts}
\usepackage{geometry}
\usepackage{placeins}
\usepackage{lscape}
\usepackage{xcolor}

\usepackage{hyperref}
\hypersetup{
    colorlinks=true,
    linkcolor=blue,
    filecolor=magenta,      
    urlcolor=cyan,
    citecolor=blue,
    pdftitle={Overleaf Example},
    pdfpagemode=FullScreen,
    }

\makeatletter
\renewcommand*\aa@pageof{, page \thepage{} of \pageref*{LastPage}}
\makeatother

\begin{document}

    \title{Statistical properties of spicules in MURaM-ChE}

    \author{Sanghita Chandra\inst{1},          
          Robert Cameron\inst{1},
          Damien Przybylski\inst{1}, 
          \and 
          Sami K. Solanki\inst{1}
          }

   \institute{Max Planck Institute for Solar System Research,
              Justus von Liebig Weg, 37077 G\"ottingen, Germany\\
              \email{chandra@mps.mpg.de}
             }

   \date{Received: October 2, 2025 / Accepted December 14, 2025}

\abstract
{Numerical simulations of the solar chromosphere have progressed towards reproducing spicules, which are transient features observed at the solar limb using spectral lines such as H$\alpha$, Ca II H\&K, or Mg II h\&k. Two types of spicules, referred to as types I and II, have been identified in observations and studied in previous numerical works. The statistics of type II spicules in 3D numerical simulations have not yet been studied.}
{We aim to compare the statistics of properties such as lengths, lifetimes, widths, heights, inclinations, and maximum velocities of self-consistently formed spicules in a MURaM-ChE simulation with observations.}
{We employ a H$\alpha$ proxy to identify fine-scale structures at the solar limb resembling spicules in a simulation of an enhanced network region. We track the evolution of 58 such features found in a 21-minute time sequence, and compare their dynamical and morphological properties with those derived from quiet-Sun observations using the Solar Optical Telescope (SOT) onboard the \textit{Hinode} mission in the  Ca II H spectral line. Previous studies have shown that spicules show very similar properties in Ca II H and H$\alpha$.}
{The spicule-like structures found in the simulation have statistical properties which are broadly consistent with those observed with \textit{Hinode}/SOT. In particular, we find evidence for the self-consistent formation of both type I and type II spicules within the simulation, even in the absence of ambipolar diffusion. We also investigate the properties of rapid blueshifted and redshifted excursions (RBEs and RREs) in the simulation in relation to the spicules.}
{}

    \authorrunning{Chandra et al.}
   \maketitle

\section{Introduction}

Spicules appear as thin, jet-like plasma structures observed at the solar limb in various chromospheric spectral lines, most prominently in  H$\alpha$ and Ca II H $\&$ K \citep{secchi_1871, De_Pontieu_2007a}. They are ubiquitous in the solar chromosphere which lies between the solar photosphere and the million degree hot solar corona. Due to their rapid evolution, occurrence rates, and energetic properties, they are considered prime candidates for facilitating mass and energy transport from the lower solar atmosphere to the corona, pumping energy into the upper solar atmosphere and potentially contributing to coronal heating (\citealp[and others]{Athay_1982, Pneuman_1978, Athay_2000, De_Pontieu_2017}). 

The earliest observational studies of spicules were limited in spatial resolution and suffered from a lack of coordinated multi-wavelength observations. Nonetheless, statistical studies were done which characterised spicules primarily in terms of their typical lifetimes (up to tens of minutes), heights (up to about 10 Mm above the solar surface) and upward velocities (20 - 50\;km/s) \citep{Beckers_1968, Beckers_1972}. The advent of high resolution space-based as well as ground-based observations such as data from the Solar Optical Telescope (SOT) onboard the \textit{Hinode} mission \citep{Tsuneta_2008}, the Swedish 1\;m Solar Telescope (SST) \citep{Scharmer_2003}, and \textit{IRIS} \citep{De_pontieu_2014}, led to improved statistics on several spicule properties (lifetimes, lengths, widths, heights, occurrence rates, etc.) observed in a range of spectral lines. High resolution observations also unveiled the presence of spicules at much higher speeds (20\;km/s - 300\;km/s) and much shorter lifetimes (tens of seconds) than had been previously reported \citep{De_Pontieu_2007a, Pereira_2012}. Additionally, these faster spicules were observed to disappear from chromospheric passbands and to reappear in hotter channels in the transition region such as Si IV (\citealp[]{Pereira_2014, Skogsrud_2015}; and others). This sparked renewed interest in identifying the role of spicules in the mass loading and energy transport to the transition region and into the solar corona.  

The faster sub-population of spicules was considered to form a separate category of spicules, so that they are now commonly classified into two categories --- type I and type II spicules. The classification is based on their lifetimes, velocities, heating signatures and trajectories \citep{De_Pontieu_2007a}. The type II spicules exhibit much higher apparent velocities, often exceeding a hundred kilometres per second. These belong to the same class of spicules that are seen to disappear from cooler passbands (Ca II H, H$\alpha$) and reappear in hotter channels (Si IV), showing transition region counterparts. Their on-disc counterparts are thought to be the rapid blueshifted and redshifted excursions (RBEs and RREs) \citep{Rouppe_van_der_Voort_2009, Sekse_2013b}. Unlike the driving mechanism of type I spicules which is by the leakage of p-mode oscillations \citep{De_Pontieu_2004}, the mechanism driving the type II spicules remains elusive. To unveil the true nature of spicules, bridging observations and theory is necessary with the help of realistic simulations.

  \begin{figure*}
   \centering
   \includegraphics[width=\textwidth]{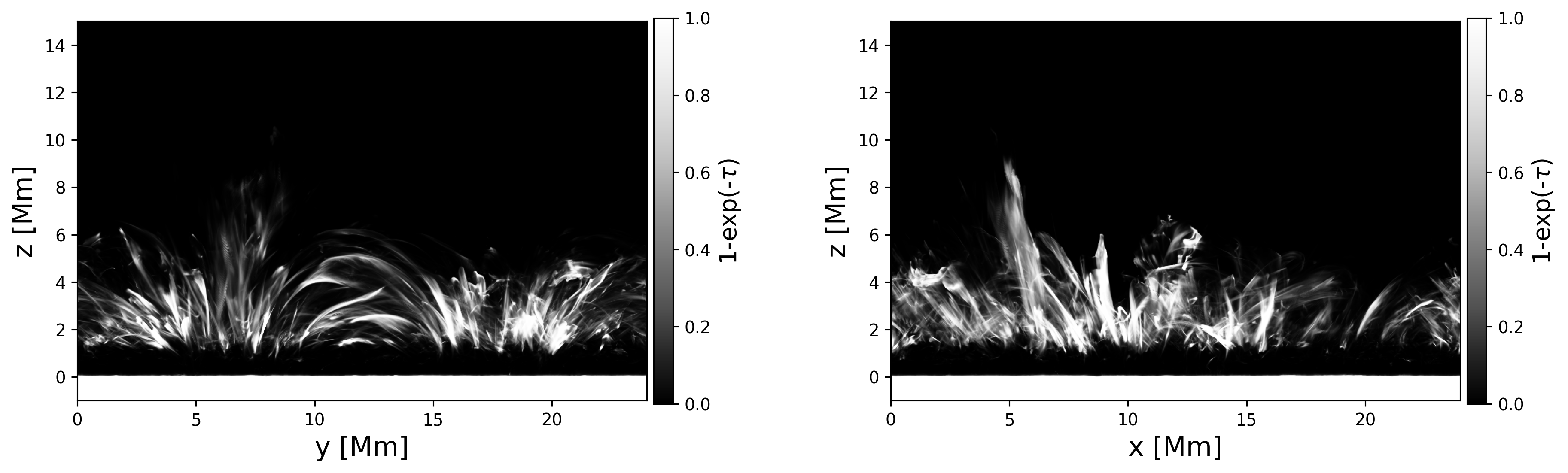}
   \caption{Spicules at the limb in the MURaM-ChE simulations, identified using the H$\alpha$ proxy from the two horizontal directions x and y respectively. The Doppler velocity ($\mathsf{v}_\mathrm{D}$) is at +37\;km/s. The photosphere is marked by 0 on the z axis. A movie showing the evolution of these spicules can be found \href{https://owncloud.gwdg.de/index.php/s/hhLnkNkrlfxoVNW}{online}.
   }
              \label{limb}%
    \end{figure*}



Realistic numerical modelling of spicules presents a range of significant challenges. Spicules form and evolve in a regime where strong magnetic fields interact with partially ionised plasma under steep temperature and density gradients. Accurately capturing these interactions requires a detailed treatment of radiative transfer in optically thick regions --- often far from local thermodynamic equilibrium (or non-LTE) \citep{Leenaarts_2012}. In addition, ionization and recombination timescales can be comparable to dynamical timescales, necessitating the use of time-dependent, non-equilibrium (NE) ionization models. Investigating the dynamics of individual features imposes further constraints on numerical resolution and temporal cadence.

Over the past two decades, several key modelling efforts have aimed to understand the formation, dynamics, and energetics of spicules. Some of the early works using 1D models include chromospheric jets driven by shock waves \citep{Heggeland_2007}. Using 2D simulations, \citet{Iijima_2015} showed the impact of coronal temperatures in determining the length of chromospheric jets. \citet{Sykora_2011} investigated the driving mechanism of a type II spicule using 3D magnetohydrodynamic (MHD) simulations, showing that the Lorentz force, in a highly inclined magnetic field topology, played a dominant role. Later, 2.5D simulations using the Bifrost code revealed that ambipolar diffusion may also be a critical factor in generating type II spicules \citep{Sykora_2017}. Recently, \citet{dey_2024} showed the generation of a forest of spicules in their 3D radiative-MHD simulations, but focused on their spinning and sheet-like morphology using synthetic Si IV emission. Synthesising chromospheric observables from such simulations, in a manner suitable for direct comparison with observations, is computationally demanding. 

The previous simulation-based works on spicules focus on individual case studies. In contrast, a statistical analysis provides a more robust connection between simulations and observations. Such studies are crucial not only for assessing the realism of simulated spicules but also for validating whether the underlying physical mechanisms are being modelled correctly.

In this paper we present a statistical study of features resembling spicules identified using a H$\alpha$ proxy \citep{Chandra_2025} in a simulation with the chromospheric extension to the MURaM code (MURaM-ChE). We focus on properties such as maximum lengths, apparent speeds, lifetimes, etc. and how these compare with their observed counterparts (e.g., \citealp[]{Pereira_2012}). The simulated spicules are generated self-consistently without imposed driving or imposed flux emergence. We do not include ambipolar diffusion in our simulation. Despite this, we find ubiquitous spicules with a range of velocities, identified with the H$\alpha$ proxy at the limb. In Sect.~\ref{model_atmosphere}, we describe the model atmosphere, followed by the H$\alpha$ proxy in Sect.~\ref{Proxy}. The results of the statistical analysis are presented in Sect.~\ref{stats}, and in Sect.~\ref{RBE_RRE_properties} we discuss the properties of RBEs and RREs in relation to spicules. Finally, we conclude with a discussion in Sect.~\ref{discussion}, and a summary in Sect.~\ref{conclusion}.


   \begin{figure*}
   \centering
   \includegraphics[width=\textwidth]{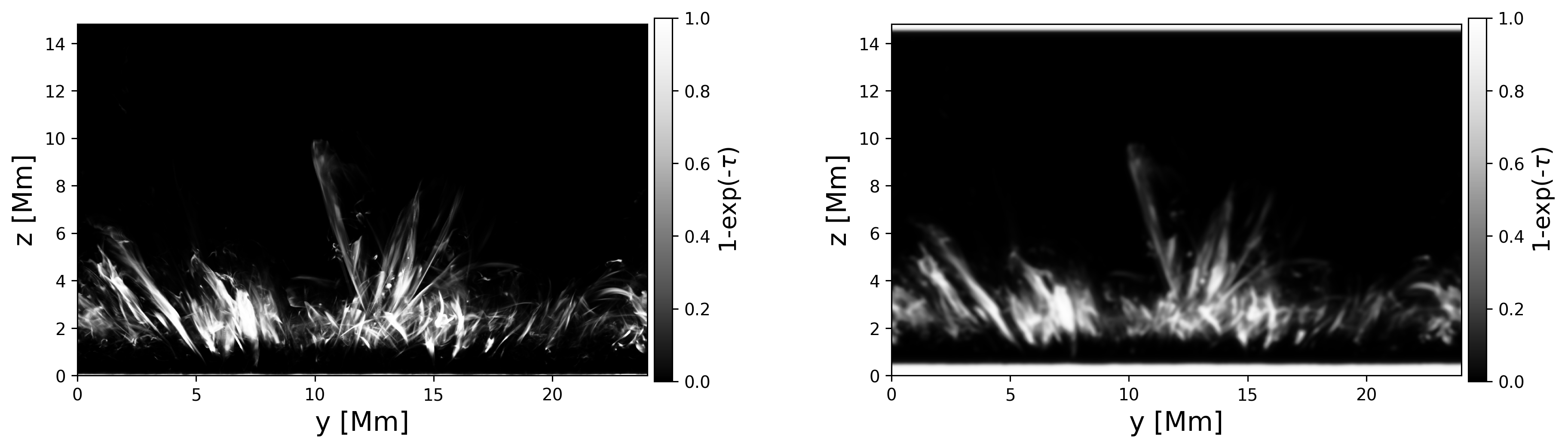}
   \caption{Comparison with \textit{Hinode}/SOT-like observation. The simulated features at a Doppler shift of -37~km/s identified with the proxy (left) is degraded with the point spread function of an ideal, spiderless 50~cm telescope (right).}
              \label{telescope}%
    \end{figure*}

   \begin{figure*}
   \centering
   \includegraphics[width=\textwidth]{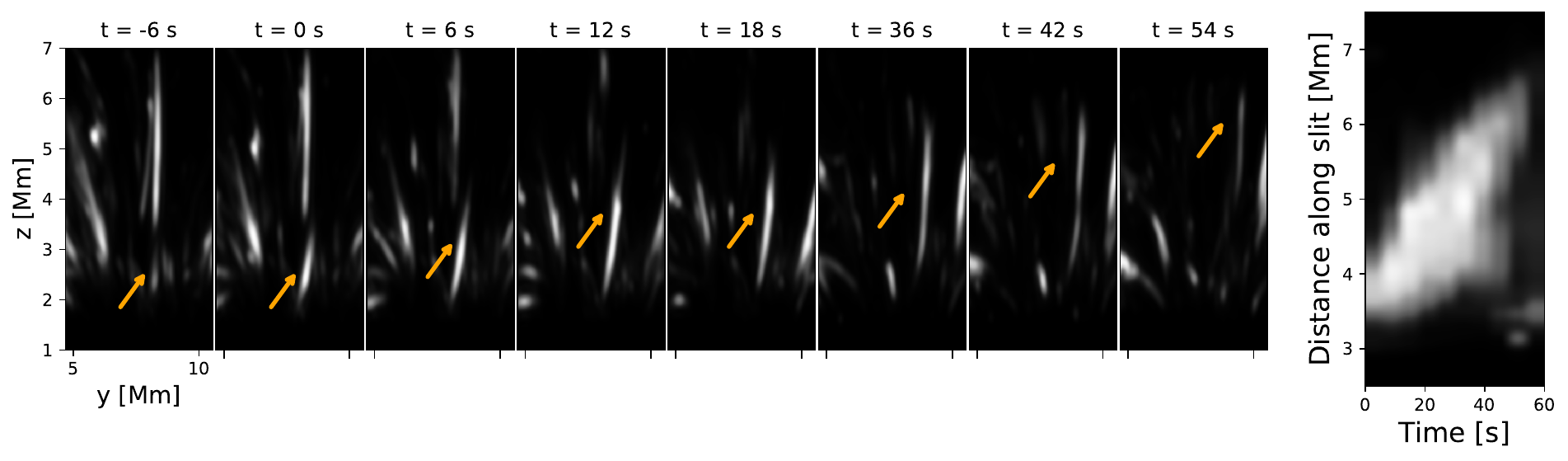}
   \caption{Evolution of a feature (after degradation to the resolution of \textit{Hinode}/SOT) resembling a type-II spicule with a maximum apparent velocity of 66\;km/s. The panels starting on the left show the evolution of our feature of interest. The orange arrows point to this feature. The time-distance plot for this feature is shown on the right. The slit for the 
   time-distance plot is placed along the axis of the feature during its evolution. The time t\,=\,0\;s marks the first prominent appearance of the feature.}
              \label{evolution}%
    \end{figure*}

   \begin{figure*}
   \centering
   \includegraphics[width=\textwidth]{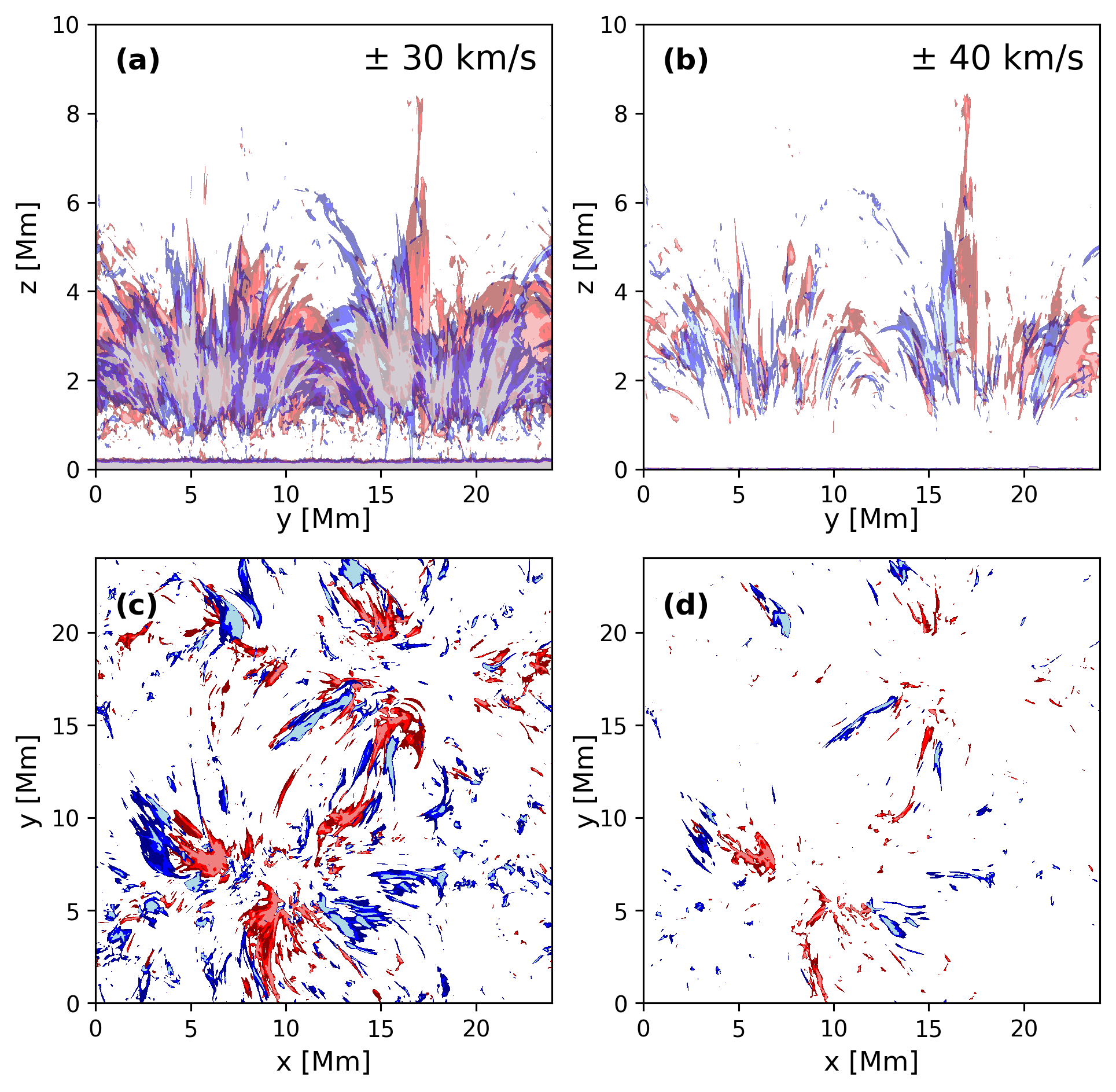}
   \caption{Spicules, RBEs and RREs seen at different $\mathsf{V}_\mathrm{{los}}$. Panels (a), (b): H$\alpha$ features viewed at the limb at Doppler velocities of +30\;km/s (blue) and -30\;km/s (red) in panel (a) and $\pm$40\;km/s in panel (b). Panels (c), (d): H$\alpha$ features on the disc at the same velocities as in panels (a), (b). The lighter shades of the colour (red/blue) indicate features with higher opacity. The evolution of the features for the entire time-series can be found \href{https://owncloud.gwdg.de/index.php/s/ugf1BIH2RB8613R}{online}.}
              \label{spicules_RBE_RRE}%
    \end{figure*}

\section{Model Atmosphere: Enhanced network model}\label{model_atmosphere}

We use the chromospheric extension, MURaM-ChE \citep{Przybylski_2022} of the MURaM radiative-MHD code \citep{Voegler_2005, Rempel_2017} to simulate the near-surface convection, photosphere, chromosphere, and corona in a 3D Cartesian setup with uniform grid spacing. MURaM-ChE incorporates a NE equation of state, solving time-dependent rate equations for hydrogen level populations --- an essential step for modelling H$\alpha$ formation \citep{Leenaarts_2007}. MURaM-ChE also includes multi-group radiative transfer with scattering \citep{Skartlien_2000, Hayek_2008}, 3D EUV back-heating of the chromosphere, and additional chromospheric and coronal radiative losses, similar to the prescription by \citet{Carlsson_2012}. We do not include ambipolar diffusion for this study.

We use a MURaM-ChE simulation of an enhanced network region (MURaM-EN) with a horiztonal extent of 24\,Mm and vertical extent of 24\,Mm spanning from 7\,Mm below to 17\,Mm above the average optical surface of $\tau_{500}$=1 (defined at z$\approx$0\,Mm). The horizontal resolution is 23.46\,km, with a higher vertical resolution at 20\,km. The simulation features a bipolar magnetic field configuration superimposed on a small-scale dynamo model \citep{Przybylski_2025} representing a solar network field patch. This magnetic configuration is similar to the public Bifrost snapshot presented in \citet{Carlsson_2016}. Boundary conditions include periodicity in the horizontal directions, an open top boundary for outflow with a potential field extrapolation for the magnetic field, and an open symmetric field bottom boundary condition \citep{Rempel_2014}. 

Our analysis is based on 206 snapshots spanning a 1280\;s time series, with a cadence of approximately 6\;s. \citet{Ondratschek_2024} provides a detailed description of the enhanced network model used in this study. Their work, which computed the Mg II h and k lines using 1.5D radiative transfer, demonstrates a closer agreement between simulated and observed line profiles (based on IRIS data) compared to previous numerical studies. The RBE structure analysed by \citet{Chandra_2025} was also formed in this simulation, exhibiting properties consistent with observations. Their work presents the identification of synthetic H$\alpha$ wing features formed on the solar disc.


    \begin{figure}
   \includegraphics[width=0.5\textwidth]{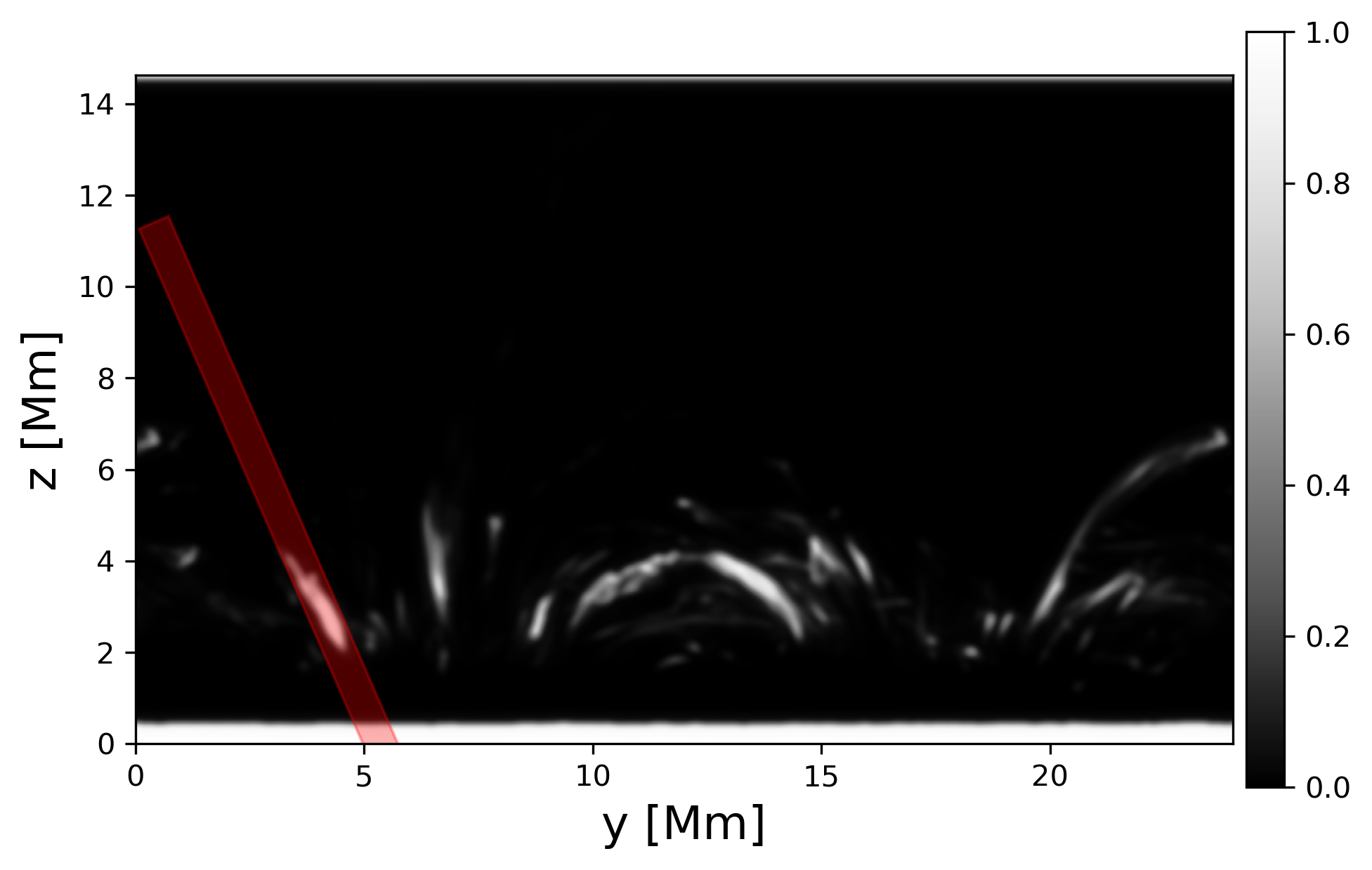}
   \caption{Tracking features at a Doppler velocity of +50\;km/s (in the same colour scale as Fig.~\ref{telescope})} by placing artificial slits over individual features. A movie showing the tracking of features can be found \href{https://owncloud.gwdg.de/index.php/s/w53dTILZp6dTIf9}{online}.
              \label{tracking}%
    \end{figure}

    \begin{figure*}
   \centering
   \includegraphics[width=\textwidth]{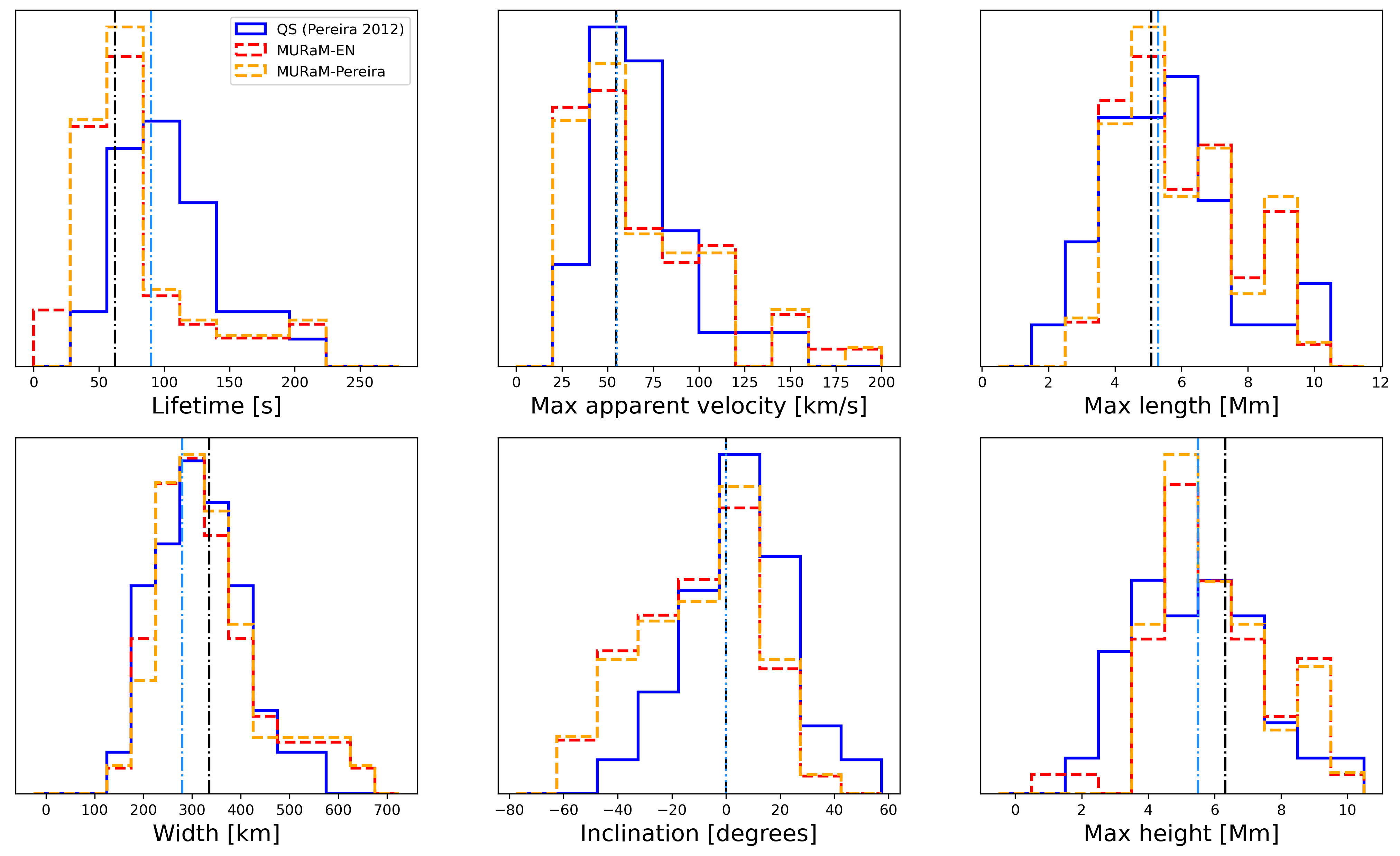}
   \caption{Statistical properties of 58 spicules in the MURaM enhanced network simulation, excluding short-lived spicules (MURaM-Pereira; see text for details) compared with 174 spicules in quiet Sun (QS) observations with the \textit{Hinode}/SOT. We also show the original distribution (MURaM-EN) without excluding short-lived spicules in red. All the distributions are normalised. The dot-dashed blue lines indicates the median value for the QS observations, while the dot-dashed black lines indicate the median value for the simulations given the lifetime criteria (MURaM-Pereira).}
              \label{histograms}%
    \end{figure*}


\section{Spicules in the H$\alpha$ proxy}\label{Proxy}

\subsection{Prescription of the proxy}

In \citet{Chandra_2025} we introduced a  H$\alpha$ proxy designed to study on-disc H$\alpha$ wing features. The  proxy is based on a photon escape probability using the NE hydrogen populations (of the n=2 and n=3 atomic levels), temperature, and line-of-sight velocity. It serves as a diagnostic to identify H$\alpha$ features formed in the simulation domain, while accounting for absorption in the H$\alpha$ spectral line.

In this study we modify the proxy to study off-limb H$\alpha$ structures where the spectral line is observed in emission. We modify the original H$\alpha$ proxy assuming that the absorbed photons are scattered along the line-of-sight. In the following, we describe the construction of the proxy for isolating off-limb H$\alpha$ features. The description is almost identical to our previous work, with the line-of-sight velocity and the direction of integration changed based on the viewing angle: from vertical to horizontal (see Eqs. \ref{line_profile} and \ref{tau}).

The line profile function ($\Phi$) assumes the form,

   \begin{align} \label{line_profile}
         & & {\Phi(\mathsf{v}_\mathrm{D}) = \frac{c}{\nu_0}\sqrt{\frac{{m}}{ {2\pi k_\mathrm{B} T}} }  \mathrm{exp}\left(-\frac{m(\mathsf{v}_\mathrm{los}-\mathsf{v}_\mathrm{D})^2}{2k_\mathrm{B}T}\right)},
   \end{align}

 \noindent  where `$\mathsf{v}_\mathrm{los}$' is the line-of-sight velocity ($\mathsf{v}_\mathrm{x}$ or \textit{$\mathsf{v}_\mathrm{y}$} in our case) in the horizontal direction. \textit{$\mathsf{v}_\mathrm{D}$} is the Doppler velocity, where positive values indicate blueshifted features, while negative values indicate redshifted features. \textit{T} is the temperature, \textit{m} is the atomic mass of hydrogen,  \textit{$k_\mathrm{B}$} is the Boltzmann constant, \textit{c} is the speed of light, and \textit{$\nu_0$} is the frequency corresponding to the rest wavelength of H$\alpha$ at the line core (in air).
\newline We compute the absorption coefficient ($\mathrm{\kappa}$) as, 

   \begin{align}\label{abs_coeff}
      & & \kappa = \frac{h \nu_0}{4 \pi} (n_2 B_{23} - n_3 B_{32}) \  \Phi(\mathsf{v}_\mathrm{D}) ,
   \end{align}

\noindent where \textit{$n_2$}, \textit{$n_3$} correspond to hydrogen populations in the (n=2) and (n=3) energy levels respectively. \textit{$B_{23}$}, \textit{$B_{32}$} are the Einstein coefficients.

\noindent The optical depth for a distance \textit{s}, is then given by

   \begin{align}\label{tau}
      & & \tau(s) = \int \kappa ds ,
   \end{align}

\noindent where \textit{ds} is the length element along the line-of-sight. For our simulation box, \textit{ds} = \textit{dx} (or \textit{dy}), and the integration is performed across the entire horizontal extent of the box. Finally, the probability of a photon traversing a distance \textit{s} without getting absorbed, or the photon escape probability, \textit{P(s)}, is 

\begin{align}
    & & P(s) = \mathrm{exp}(-\tau).
\end{align}

  This formulation describes the absorption of photons similar to \citet{Chandra_2025}. The absorbed photons are then assumed to be scattered along the line-of-sight or horizontal directions. At the limb the features opaque in H$\alpha$ will appear in emission, as they scatter the radiation emitted from the solar photosphere along the line-of-sight. This enables the identification of off-limb H$\alpha$ structures. For the proxy at the limb we use the quantity (\textit{1-P(s)}), such that 1 denotes a feature and 0 corresponds to no feature. With this prescription, we find ubiquitous features at the limb in the simulation (see Fig.\;\ref{limb} and the associated movie).

  We note that in \citet{Chandra_2025}, we used the 3D absorption coefficient or opacity (Eq. \ref{abs_coeff}) information to show the contribution of the RBE in vertical slices. However, the observable we developed in this work involves integrating the opacity along the horizontal direction as the line-of-sight (see Eq. \ref{tau}). This mimics the emitted radiation and helps to capture off-limb H$\alpha$ structures as they appear in observations. However, we assume that the radiation seen by the observer primarily depends on the line-of-sight velocity and the hydrogen populations. In reality, the atoms are not uniformly lit from below and can move at high velocities relative to the photosphere, resulting in Doppler brightening effects. This can influence how spicules appear in our synthetic observable and would need to be accounted for to obtain a more realistic treatment.

\subsection{Synthetic observations degraded to the resolution of \textit{Hinode}/SOT}

To simulate telescope imaging, we model the point spread function (PSF) using an Airy disc, which describes diffraction from a circular aperture. The resolution limit is determined by the first minimum of this pattern. We generate a 2D PSF based on the telescope’s aperture (taken to be 50\;cm, which corresponds to the aperture of \textit{Hinode}/SOT) and the observed wavelength of H$\alpha$ in air. Finally, the normalised PSF is convolved with the simulated image to replicate how a Hinode-like telescope would capture the spicules. This is shown in Fig. \ref{telescope}. We note that because the telescope spiders were not explicitly modelled, the image quality of actual \textit{Hinode} observations may be slightly lower than the images shown here. \cite{Gunar_2019} modelled the instrumental effects of the \textit{Hinode}/SOT NFI for H$\alpha$ by accounting for the Lyot-filter transmission profile, adding realistic noise, and applying the instrument’s PSF to synthetic prominence images. They found that fine structures remain visible even after degradation. Similarly, the degraded spicules in our analysis largely preserve the appearance of the original ones.

The simulated spicules exhibit morphological similarities to those observed in solar limb observations (e.g., \citealp[]{Zhang_2012,Pereira_2012, Pereira_2016}), although these studies mainly used the Ca II H line for observations. The evolution of an example spicule at a Doppler shift of +50\,km/s is shown in Fig.\,\ref{evolution}. It is seen to disappear from the passband and exhibits (mostly) a linear trajectory, with a maximum velocity of 66\,km/s. This feature resembles a type II spicule. For the statistical analysis presented below, we use such spatially degraded images of the simulated spicules to facilitate a more direct comparison with observations.

\section{Statistical properties of spicules}\label{stats}

Using the H$\alpha$ proxy, we detect a large number of features at lower Doppler velocities (20\;–\;40\;km/s). As we move further into the wings of the H$\alpha$ proxy (average escape probability profile), corresponding to higher Doppler shifts of 40\;–\;60\;km/s, the number of detected features decreases significantly (Fig. 
 \ref{spicules_RBE_RRE}). This trend is evident both in off-limb spicules and in on-disc RBEs \citep{Chandra_2025} and RREs within our simulations. The velocity distribution of these features shows that higher line-of-sight velocities are less common, indicating that fewer structures exhibit rapid motion along the line of sight. This behaviour is consistent with observations of spicules reported by \citet{Pereira_2016}.

 We studied 58 individual spicules at a Doppler velocity of $\pm$\,50\;km/s. All of the features were detected and studied manually. An example of feature tracking is shown in Fig. \ref{tracking} and the associated movie. We place artificial slits over features to track them. These slits are fixed in time for a given feature, and were placed in a way to capture the spicule at its maximum length. We found both type I and type II spicules in the simulations. We focused on the type II spicules by choice to later understand the driving mechanism behind their formation. We show the distribution of six spicule properties (lifetime, apparent speeds, maximum length, width, inclination, and maximum height) for the identified spicules in MURaM-EN. We then compare the distribution of these properties
 with those observed for quiet-Sun spicules (174 spicules in total) in the study presented by \citet{Pereira_2012}. The corresponding histogram distribution can be seen in Fig. \ref{histograms}. 
 Tables~\ref{table:1} and \ref{table:2} further summarise the key statistical properties (range, mean, and standard deviation) for the observed and simulated spicules.


\begin{table}
\caption{Statistical properties of spicules}\label{table:1}
\centering
\begin{tabular}{ |c|c|c| } 
 \hline
 Property & MURaM-Pereira & Observations (QS) \\
 & (min - max)&(Pereira et al. 2012) \\

 \hline
 \hline
 Lifetime [s] & 40 - 150 & 50 - 150 \\
 Max velocity [km/s] & 20 - 230 & 18 - 160 \\ 
 Max length [Mm] & 2.3 - 8.9 & 1.5 - 10 \\
 Width [km] & 200 - 500 & 200 - 450 \\
 Inclination [$^{\circ}$] & -46 - 56 & -40 - 50 \\
 Max height [Mm] & 1 - 10 & 2 - 10 \\ 
 \hline
\end{tabular}\
\vspace{0.1cm}
\tablefoot{Statistical properties of 30 spicules in the simulations versus 174 spicules in quiet-Sun (QS) observations with \textit{Hinode}/SOT \citep{Pereira_2012}. The lifetimes and widths of the observations are reported as the values where the majority of spicules (mode of the distribution) lie.}
\end{table}



\begin{table}
\caption{Mean properties of spicules}\label{table:2}
\centering
\begin{tabular}{ |c|c|c| } 
 \hline
 Property & MURaM-Pereira & Observations (QS) \\
 & (mean/std dev)&(Pereira et al. 2012) \\

 \hline
 \hline
 Lifetime [s] & 80/45 & 108/49 \\
 Max velocity [km/s] & 70/42 & 61/23 \\ 
 Max length [Mm] & 5.49/1.72 & 5.48/1.79 \\
 Width [km] & 365/106 & 304/78 \\
 abs(Inclination) [$^{\circ}$] & 17.4/12.7 & 12.7/9.8 \\
 Max height [Mm] & 6.66/1.76 & 5.48/1.70 \\ 
 \hline
\end{tabular}\
\vspace{0.1cm}
\tablefoot{Mean spicule properties comparing the quiet Sun observations with the enhanced network simulation, imposing the lifetime criterion (MURaM-Pereira). The first value for each property indicates the mean and the second value shows the standard deviation.}
\end{table}


 \noindent \textit{Lifetime:} The lifetimes of the features were determined through visual inspection at the high Doppler shifts ($\pm$\,50\;km/s). The choice of high Doppler velocity minimises the likelihood of superposition with unrelated structures. We chose very dynamic features resembling type II spicules which are short-lived, thereby biasing our analysis towards shorter lifetimes (MURaM-EN, Fig.~\ref{histograms}). \cite{Pereira_2012} used only the linear spicules (174 out of 177) for their statistics. Moreover, to reduce false positives in their detection technique they reject spicules that last less than 32-45 s. Following this, we impose a lifetime criterion where we discard spicules lasting less than 32\,s, and prepare an additional distribution (MURaM-Pereira) shown in orange in Fig.~\ref{histograms}. This distribution is also shown for the other spicule properties. We analyse spicules in a time series of only 21 minutes, which limits the variety of spicules we find, compared to the longer datasets used by \cite{Pereira_2012}. We find that the mean lifetime of our identified spicules is at 80 seconds (see Table \ref{table:2}). This is less than the mean value for observed quiet Sun spicules at 108 seconds. This disagreement may have to do with our simulation setup in addition to our selection bias in choosing the very dynamic spicules at a fixed high Doppler shift of $\pm$\,50\,km/s.

\noindent \textit{Maximum apparent velocity:} To estimate the apparent velocities of the spicules, we place artificial slits along their lengths and track their evolution over time until they disappear. Using our proxy, which approximates the emergent H$\alpha$ intensity, we construct space-time diagrams and extract the maximum apparent velocity from the slopes of the feature tracks. In some cases, we observe sharp changes in slope, possibly indicating more complex dynamics such as torsional or swirling motions \citep{B_de_pontieu_2012, Sekse_2013a}. For such cases we report the faster apparent velocity. Most spicule-like features exhibit apparent velocities in the range of 25\;km/s\;–\;100\;km/s, although we identify one exceptionally fast event reaching approximately 230\;km/s.

 \noindent \textit{Maximum length:} We estimate the maximum length of spicules by tracking their evolution along artificial slits placed parallel to their apparent orientation or the spicule axis. The length is measured from the proxy maps based on where the spicule is seen to end, which is not always at the limb. To remain consistent with the assumption that all spicules are connected to the limb, adopted by many observational studies including \citet{Pereira_2012}, we added $\sim$\,1\,Mm to our computed value to connect the spicules to the limb. The maximum observed length for the spicules we study reaches approximately 10\;Mm. However, a key limitation of this method is that many spicules exhibit curvature, while our measurement assumes a linear (projected) path. This approximation introduces a bias toward underestimating the true lengths and excludes loop-like or highly curved features from our analysis. Within these constraints, we find that most spicule-like structures reach maximum lengths in the range of 4\;–\;7\;Mm. 

 \noindent \textit{Width:} The width is defined as the maximum extent to which a spicule broadens perpendicular to its axis during its evolution. To estimate this, we first choose slit widths that encompass the full lateral extent of each spicule (see Fig. \ref{tracking}). At the time when the spicule reaches its peak intensity, we extract a cross-sectional intensity profile perpendicular to the slit axis. This profile is then fitted with a Gaussian function, and the full width at half maximum (FWHM) of the fit is reported as the spicule width. Most of the spicules we analysed are narrow structures, with widths ranging from approximately 200 to 500\;km. Unlike in \citet{Pereira_2012}, we do not choose a constant height at which the width is measured. Rather, we measure the width where the spicule reaches maximum intensity (with the proxy) along its axis. We find a tail in the distribution showing broader features, when compared with observations. This is consistent with observational studies reporting larger widths for H$\alpha$ spicules (e.g. \citealp{Pasachoff_2009})

 \noindent \textit{Inclination:} The inclination is measured by taking the angle between the artificial slit for a given spicule and the normal to the surface (or limb), i.e., the vertical. We record the average inclination for a given spicule over its lifetime. Most of our spicules are roughly vertical, with an inclination at 0$^{\circ}$.

 \noindent \textit{Maximum height:} The maximum height reached by each spicule is measured relative to the average solar surface. 
 Since our enhanced network model primarily consists of low-lying magnetic loops, we do not expect spicules to reach the same heights as those typically observed in quiet Sun or coronal hole regions. Additionally, there may be an observational bias favouring the detection of taller spicules, as these rise above the dense "spicule forest" and are more easily distinguished. At the same time we include spicules, the faint tops of which can be identified using the proxy -- a criterion that introduces subjectivity in observations which have already undergone image processing such as radial filtering and embossing. This can lead to identifying taller spicules compared to their observed counterparts. Despite these considerations, we find a good overall agreement between the heights of simulated and observed spicules. Most of the spicules in our study reach heights of approximately 4\;–\;7\;Mm above the solar surface. We also find one feature that reaches about 10\;Mm in height, which can be seen as a faint structure in Fig.~\ref{telescope}.

 Overall, we find a reasonably good match between the observed and simulated spicules in terms of their statistical properties. The median values of the spicule properties are indicated in Fig.~\ref{histograms}, while the mean values and standard deviation are shown in Table~\ref{table:2}. The mean and median values of the maximum velocity, lengths, widths, inclination and maximum heights have good agreement. This is not true for the lifetime of the detected spicules. Our simplified detection techniques could contribute to the differences in the distribution of properties. The use of straight slits underestimates the length of spicules with some curvature, apart from excluding highly curved spicules from the analysis altogether. Additionally, our emphasis on dynamic type II spicules led to the detection of more features with shorter lifetimes.

 Even without including ambipolar diffusion \citep{Sykora_2017} in the simulation setup, we obtain type II spicules with apparent speeds reaching $\sim$ 230 km/s. The maximum heights of our simulated spicules reach 10\,Mm, consistent with observed characteristics of spicule formation and dynamics.


    \begin{figure*}
   \centering
   \includegraphics[width=\textwidth]{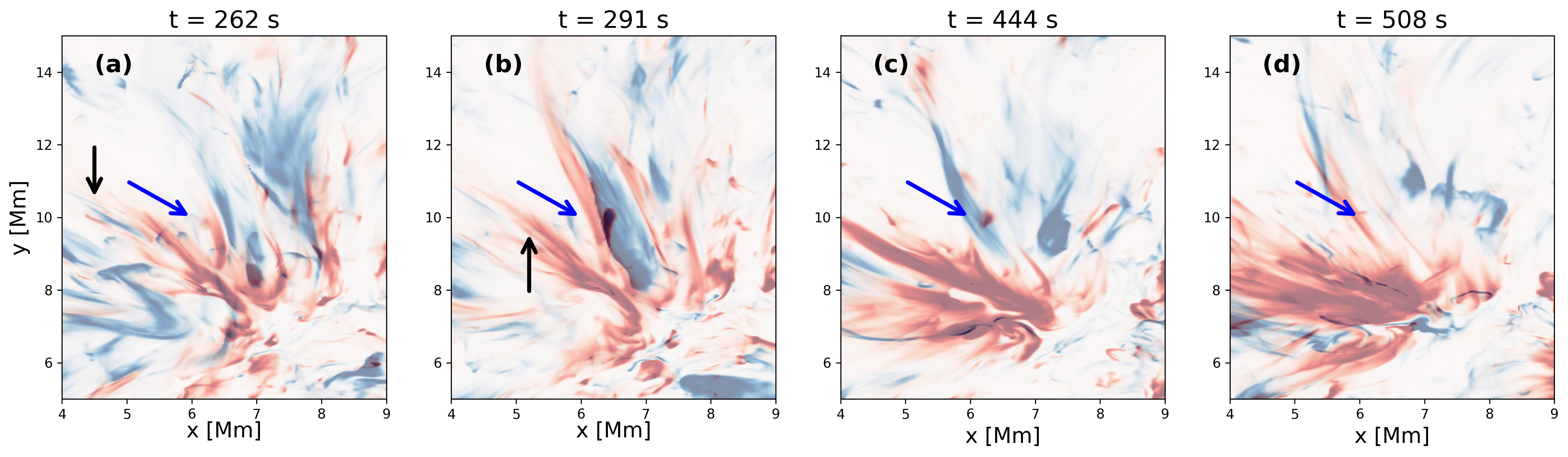}
   \caption{Evolution of overlapping RBE (blue) and RRE (red) structures and their multi-threaded nature at the original resolution of the simulation. The blue arrows point to an RBE structure evolving close to an RRE structure. The black arrows indicate the multi-threaded nature of some of these structures.}
              \label{RBE_RRE_structure}%
    \end{figure*}


\section{Relationship of the properties of spicules with RBEs/RREs in MURaM-EN}\label{RBE_RRE_properties}

We also examine on-disc features using the H$\alpha$ proxy from a top-down perspective (same as the approach used in \citealp[]{Chandra_2025}). This enables us to investigate the hypothesis that RBEs and RREs represent the on-disc counterparts of the faster type II spicules. Furthermore, we can independently assess their statistical properties. From the movie associated with Fig. \ref{spicules_RBE_RRE}, several qualitative properties can be inferred that are consistent with observational signatures of RBEs and RREs. The so called downflowing RREs, first identified in high-resolution Swedish 1-m Solar Telescope (SST) observations (e.g., \citealp{Bose_2021}), are characterised by plane-of-sky motion directed toward magnetic network concentrations. This behaviour contrasts with traditional RBEs and RREs, which typically exhibit outward motion away from regions of enhanced magnetic field. In our simulation, we observe both types of RRE structures. For instance, some representative downflowing RRE structures can be seen in the movie associated with panel (c) of Fig. \ref{spicules_RBE_RRE} near coordinates (x, y) = (11\;Mm, 5\;Mm).

We also identify regions where an RRE appears shortly after the disappearance of an RBE at the same location, or where both features are co-spatial. This is shown in Fig. \ref{RBE_RRE_structure}, where an RBE overlaps with an RRE (panel b) and eventually disappears while an RRE takes its place (panel d). This suggests the presence of cyclical mass flows, where RBEs may be responsible for upward mass transport, followed by a downward return flow traced by RREs. Alternatively, such behaviour could arise from torsional or Alfv\'enic wave motions that produce alternating Doppler shifts \citep{Sekse_2013a}.

The multi-threaded nature of spicules and their on-disc counterparts, previously discussed in numerous observational and numerical studies \citep{Rouppe_van_der_Voort_2009, Pereira_2014, Bose_2021}, is also reflected in our simulation. This is evident in the complex spectral signatures of RBEs and RREs, where features appear and disappear at different Doppler velocities, suggesting a plasma composed of multiple components moving at different speeds. Upon closer inspection of individual events, we also detect finer-scale structures embedded within the main features (indicated by the black arrows in panels a, b in Fig. \ref{RBE_RRE_structure}), moving coherently, consistent with the interpretation of spicules as composed of dynamically evolving threads.

\section{Discussion}\label{discussion}

The statistical analysis of spicules in this study is primarily based on their morphological properties. \citet{Pereira_2013} utilised data from the narrowband filter imager (NFI) of SOT \citep{McIntosh_2008} to generate synthetic H$\alpha$ filtergrams at Doppler shifts of $\pm$37~km/s. These synthetic filtergrams were then compared with the observations of the Ca II H line Broadband Filter Imager (BFI) onboard SOT, which correspond to the same dataset as used by \citet{Pereira_2012}. The comparison revealed a good resemblance between spicules identified in H$\alpha$ and those observed in Ca II H (see Fig.~3 in \citealp{Pereira_2013}). Based on this agreement, the authors concluded that it is reasonable to compare spicule statistics derived from the two spectral lines using morphological criteria. Building on this foundation, we compared the statistical properties of spicules identified in the H$\alpha$ proxy with those detected in Hinode/SOT Ca II H observations.

Despite the broad similarity in the appearance of spicules in Ca II H and H$\alpha$ diagnostics, several differences remain that can influence statistical measurements. First, Ca II H filtergrams from SOT/BFI are relatively insensitive to Doppler motions, whereas our H$\alpha$ proxy is not. Large line-of-sight velocity excursions can shift a spicule in and out of the fixed H$\alpha$ passband near $\pm$\,50 km/s, artificially shortening its measured lifetime. This is an inherent limitation of our choice of a relatively large Doppler offset although it is useful for mitigating line-of-sight superposition. Second, spicule tips identified in Ca II H often appear fainter than in H$\alpha$ because of the lower chromospheric opacity of the Ca line and the broader SOT/BFI filter, which mixes photospheric and low-chromospheric contributions. The photospheric contribution in the BFI makes it easier to trace down spicules to the limb in Ca II H, whereas H$\alpha$ line-core images exhibit a chromospheric haze just above the limb. Conversely, the upper parts of spicules can appear brighter in H$\alpha$, sometimes producing slightly larger apparent heights. Nonetheless, high-resolution multi-line observations show that the overall morphology and evolution of spicules are broadly similar in Ca II H and H$\alpha$ \citep{Pereira_2016}.

We use relatively large Doppler shifts of $\pm50$\,km/s for our analysis of spicules. This exceeds the Doppler shifts of $\pm37$\,km/s used by \citet{Pereira_2013} in their comparison of H$\alpha$ and Ca II H spicules. However, \citet{Pereira_2016} analysed the behaviour of H$\alpha$ spicules observed with SST/CRISP over a broader range of Doppler shifts, extending up to $\pm50$\,km/s (see their Fig.~3). The spicules in our simulations appear largely similar to those shown in the latter study.

The enhanced network model used in our simulations contains numerous low-lying magnetic loops. Several of the larger features identified using the H$\alpha$ proxy form loop-like structures that reach heights of approximately 2\;–\;4\;Mm. These loops are more prominent when viewed along the x-direction (see Fig. \ref{limb}), and their on-disc counterparts are rooted in the enhanced network regions (see Fig.~\ref{spicules_RBE_RRE}). The bipolar structure of this enhanced network patch can be found in Fig.~1 of \cite{Chandra_2025}. The spatial association of most RBEs and RREs clustering around magnetic network boundaries aligns with observations (Fig.~\ref{spicules_RBE_RRE}).

We find co-spatial occurrence of RBEs and RREs (see Fig. \ref{RBE_RRE_structure}), including examples of downflowing RREs (movie associated with Fig. \ref{spicules_RBE_RRE}). This suggests the presence of bi-directional flows or return motions, consistent with previous observational reports \citep{Bose_2021, Chaurasiya_2024}.

There are some differences in the statistical properties of the spicule-like features identified in the MURaM-EN simulation when compared with spicules observed using the H$\alpha$ and Ca II H lines. Our selection criteria, which emphasised dynamic and fast-evolving features, especially those resembling type II spicules, resulted in the detection of more short-lived and relatively straight features with minimal curvature. This introduces a bias toward shorter lifetimes compared to quiet Sun observations (e.g., \citealp[]{Pereira_2012}), where a broader range of spicule dynamics is captured. The spicules detected in their work also suffered from selection effects and errors (see Sect. 3.3, \citealp[]{Pereira_2012}) such as discarding short spicules with typically short lifetimes. This further accounts for some of the discrepancy we find in the present study.

The maximum apparent velocities span a wide range, with some of the features reaching up to 230\;km/s. This is broadly consistent with observations, where the fastest spicules are typically found in coronal holes, followed by those in the quiet Sun \citep{Pereira_2012, Tei_2025}. However, features exceeding 100\;km/s in chromospheric diagnostic lines remain relatively rare both in simulations and observations.

The maximum simulated spicule lengths cluster around 5–6\,Mm, consistent with observed values. Observations, however, suffer from selection biases: longer spicules are more likely to be detected because they extend above the dense “forest of spicules,” while the faint tips of spicules in Ca II H data can cause their lengths to be underestimated. In our measurements, the spicule length is computed from the base of each structure along the artificial slit (see Fig.~\ref{tracking}), rather than from the limb surface, to reduce projection effects and improve tracking. To remain consistent with most observational studies, we add an approximate 1\,Mm to account for the portion extending into the underlying “spicule forest” \citep[e.g.][]{Skogsrud_2015}. Similarly, \citet{Pereira_2012} assumed that all spicules connect to the limb surface and therefore added an inferred length for the segment not directly visible in the images. On-disc RBEs and RREs identified with the same proxy show good agreement with observed lengths.
We also note that both the detectability and measured length of spicules depend on the Doppler shift at which they are observed. In this study, we identify features only at Doppler velocities of $\pm$\,50\,km/s to minimise confusion from overlapping structures.

When studying the spicules at the original resolution of the simulation, many of them have widths near the resolution limit of the Hinode telescope ($\sim$\,150\;km). Thus using the simulation allows us to resolve narrow structures that are otherwise difficult to identify observationally. At the same time, we also detect broader features during the evolution of the spicules, which could arise either from single extended structures or from merging events. Multiple spicules could form in close proximity and appear as a single structure. This is especially prominent at lower Doppler shifts, where overlapping structures are more prevalent.

Inclinations measured in the simulations are generally consistent with those observed. Due to our use of a high Doppler velocity ($\pm$\,50\;km/s) to isolate individual features, the simulated limb regions are relatively free of overlapping structures, the so-called forest of spicules. In these cleaner regions, we also detect more highly inclined, low-lying features --- sometimes more inclined than typically observed spicules. Highly inclined spicules often appear shorter due to projection effects and can become obscured by overlapping structures, while more vertical ones are easier to isolate and compare with observations.

The maximum heights reached by spicules in our simulations 
agree reasonably well with their observed counterparts. The majority of long spicules are found along the few open field lines connected to the bipolar magnetic patches. In contrast, the bulk of the magnetic field forms a dense network of closed loops that extend only to heights of 2~–~6~Mm. These closed loops have high curvature and are excluded from our study.

Finally, while some spicules in our simulations show distinct transverse motions (see the spicule in Fig. \ref{evolution}), this behaviour is not ubiquitous. A subset of spicules exhibits complex dynamics that go beyond simple lateral displacements, including twisting or swaying behaviour. However, due to the limited number of clear cases, we have not conducted a statistical analysis of transverse motions in this study.

\section{Conclusion}\label{conclusion}

Overall, we find a reasonable agreement in a range of parameters like the maximum height, length, width, maximum velocity, and inclination, and some discrepancy in the lifetime between the observed and simulated spicules. The discrepancy may have to do with limitations in our setup and simplifications in our measurement techniques in choosing the spicules. Nonetheless, the consistency in the statistical distribution of several observed properties of spicules and their characteristic ranges lends confidence to our simulation results. In addition, we have the full three-dimensional information of these structures formed in the chromosphere with our simulations. This can be used to further investigate the true velocities and morphology of spicules. This is a topic addressed in an upcoming paper.

We use the work by \citet{Pereira_2012} on quantifying spicules using \textit{Hinode}/SOT observations as a reference to compare our simulated spicules in an enhanced network region using MURaM-ChE. We use a H$\alpha$ proxy to detect the spicules in the simulation, unlike in the observations where the Ca II H line was used. Nevertheless, we expect the statistical properties to be similar based on earlier observational works comparing spicules in Ca lines and H$\alpha$ \citep{Rouppe_van_der_Voort_2009, Sekse_2012, Pereira_2013, Pereira_2016}. However, we cannot rule out that some of the discrepancy between the simulated and observed spicules arises from the different spectral lines considered.

Although we do not include ambipolar diffusion \citep{Sykora_2017} in the simulation setup, we obtain type II spicules with fast apparent speeds (up to $\sim$ 230\,km/s). The maximum heights of the spicules reach up to 10 Mm, similar to those observed. Our results suggest that our simulation framework captures many of the essential physical processes governing spicule formation and dynamics, and provides a valuable tool for interpreting and complementing high-resolution solar observations.

\begin{acknowledgements}
       This work was carried out in the framework of the International Max Planck Research School (IMPRS) at the Technical University of Braunschweig. We are grateful for the computational resources provided by the Raven and Viper supercomputer systems of the Max Planck Computing and Data Facility (MPCDF) in Garching, Germany. This research has received financial support from the European Research Council (ERC) No. 101097844 (WINSUN). This work was supported by the Deutsches Zentrum f{\"u}r Luft und Raumfahrt (DLR; German Aerospace Center) by grant DLR-FKZ 50OU2201. 
\end{acknowledgements}

\bibliographystyle{aa} 
\bibliography{bibfile} 

\appendix
\section{Spicules seen in the time-distance plots}

    The time distance plots for some example spicules are shown below. All plots are shown on identical scales.

   \begin{figure}[ht!]
   \centering
   \includegraphics[width=\textwidth]{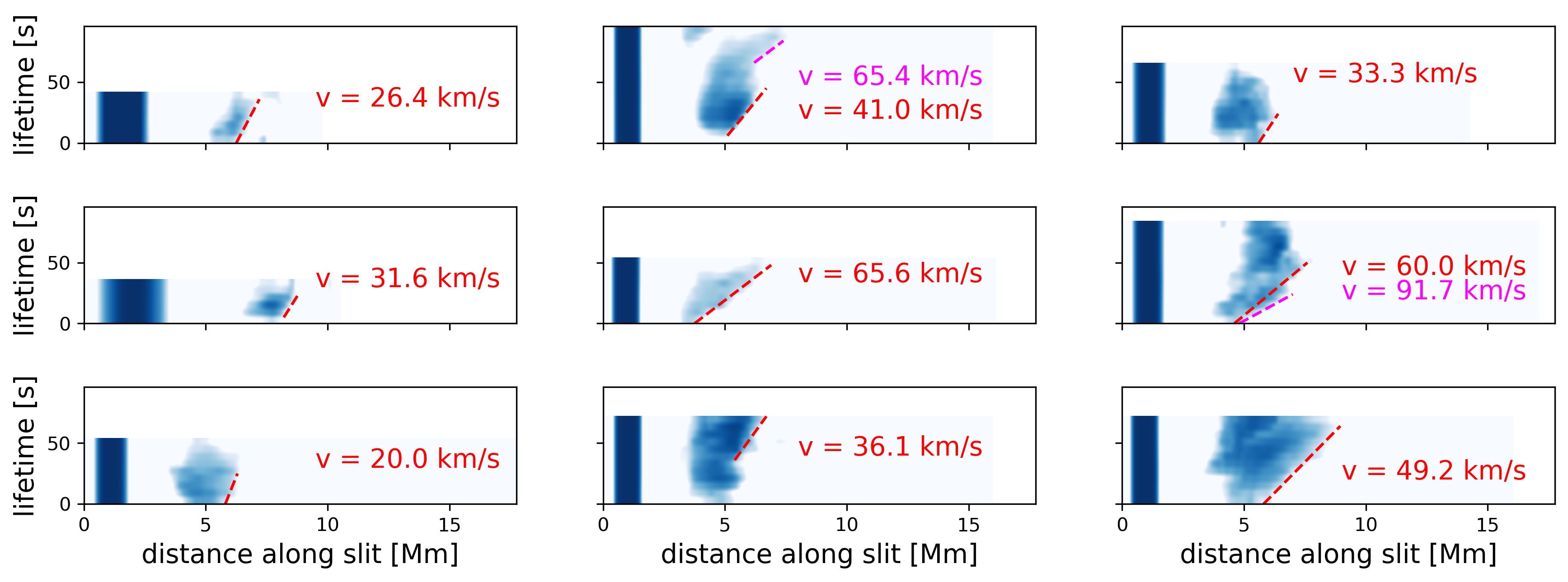}
   \caption{Time-distance plots for the spicules seen from the x direction, i.e. with the line of sight parallel to the x-axis.}
              \label{TD_telescope_sharp}%
    \end{figure}

   \begin{figure}[ht!]
   \centering
   \includegraphics[width=\textwidth]{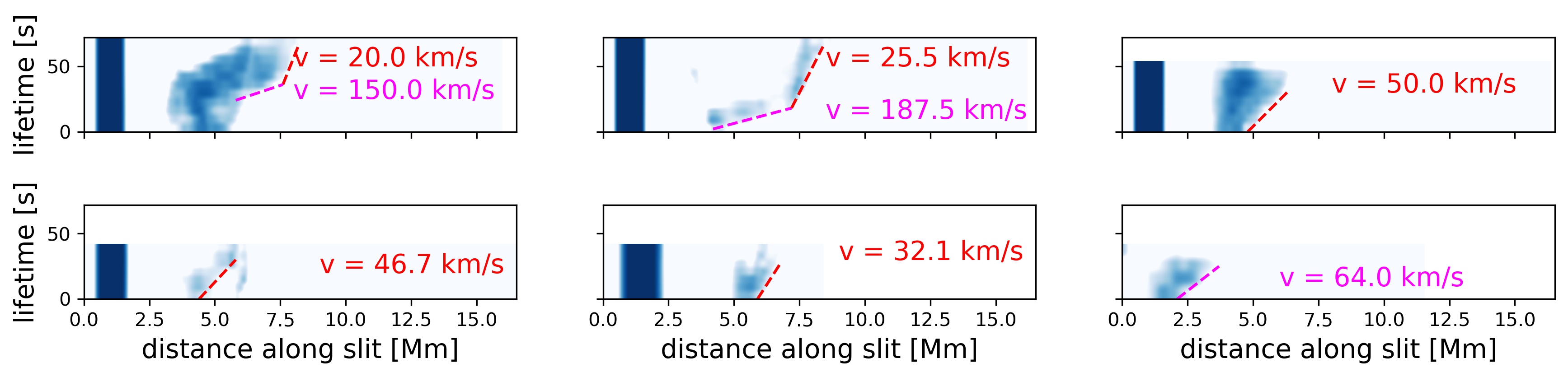}
   \caption{Time-distance plots for the spicules seen from the y direction, i.e. with the line of sight parallel to the y-axis.}
              \label{TDy_telescope_sharp}%
    \end{figure}
\end{document}